# Training of Instrumentalists and Development of New Technologies on SOFIA

Edwin F. Erickson, Louis J. Allamandola, Jean-Paul Baluteau, Eric E. Becklin, Gordon Bjoraker, Michael Burton, Lawrence J. Caroff, Cecilia Ceccarelli, Edward B. Churchwell, Dan P. Clemens, Martin Cohen, Dale P. Cruikshank, Harriet L. Dinerstein, Edward W. Dunham, Giovanni G. Fazio, Ian Gatley, Robert D. Gehrz, Reinhard Genzel, Paul Graf, Matthew A. Greenhouse, Doyal A. Harper, Paul M. Harvey, Martin Harwit, Roger H. Hildebrand, David J. Hollenbach, Adair P. Lane, Harold P. Larson, Steven D. Lord, Suzanne Madden, Gary J. Melnick, David A. Neufeld, Catherine B. Olkin, Christopher C. Packham, Thomas L. Roellig, Hans-Peter Roeser, Scott A. Sandford, Kristen Sellgren, Janet P. Simpson, John W. V. Storey, Charles M. Telesco, Alexander G. G. M. Tielens, Alan T. Tokunaga, Charles H. Townes, Christopher K. Walker, Michael W. Werner, Stanley E. Whitcomb, Juergen Wolf, Charles E. Woodward, Erick T. Young, Jonas Zmuidzinas

## *Introduction*

The Astronomy and Astrophysics 2010 Decadal Survey (Astro2010)[1] Committee has requested white papers related to the State of the Profession[2]. In response, this paper is submitted to emphasize the potential of the Stratospheric Observatory for Infrared Astronomy (SOFIA) to contribute to the training of instrumentalists and observers, and to related technology developments. This potential goes beyond the primary mission of SOFIA, which is to carry out unique, high priority astronomical research.

SOFIA is a Boeing 747SP aircraft with a 2.5 meter telescope[3]. It will enable astronomical observations anywhere, any time, and at most wavelengths between 0.3 µm and 1.6 mm not accessible from ground-based observatories. These attributes, accruing from SOFIA's mobility and flight altitude, guarantee a wealth of scientific return. Its instrument teams (nine in the first generation) and guest investigators will do suborbital astronomy in a shirt-sleeve environment. The project will invest $10M per year in science instrument development over a lifetime of 20 years. This, frequent flight opportunities, and operation that enables rapid changes of science instruments and hands-on in-flight access to the instruments, assure a unique and extensive potential - both for training young instrumentalists and for encouraging and deploying nascent technologies. Novel instruments covering optical, infrared, and submillimeter bands can be developed for and tested on SOFIA by their developers (including apprentices) for their own and guests' observations, to validate technologies and maximize observational effectiveness.

Although SOFIA's breadth in wavelength coverage, instrument capability, and observing flexibility guarantee that it will make major contributions in important areas of astrophysics, SOFIA's contributions to science are not the subject of this current white paper.

## *Airborne Astronomy Heritage*

SOFIA will promote the advancement of needed technologies and grow the competencies of the next generation with relevant instrumentation. Our confidence in this potential is based on experience from the airborne astronomy program that operated at NASA Ames Research Center from 1965 to 1995, and in particular on the 21 years of achievement of the Kuiper Airborne Observatory (KAO). Its many accomplishments – for example early evidence for hot stars and a black hole at the Galactic Center (based on far-infrared spectroscopic observations made when based in Honolulu and Christchurch, New Zealand), and discovery of the rings of Uranus (based on optical observations of a stellar occultation made when based in Perth, Australia) – attest to the effectiveness of the KAO *modus operandi*.

A primary factor in the scientific success of the KAO was the vigorous and productive science instrument-development program it spawned in the science community. Sixteen of the instruments existing in 1995, listed in Table 1, exhibit the wide range of technologies made available by the instrument teams *for observations not possible from ground-based sites*.

Table 1. Kuiper Airborne Observatory Focal Plane Instruments Existing in 1995

| Principal Investigator/ Affiliation | Instrument Type | Wavelength Range (µm) | Spectral/Spatial Channels | Spectral Resolution |
|---|---|---|---|---|
| A. Betz / U. Colorado | Heterodyne Spectrometer | 60-400 | 512/1 | $\delta\nu = 3$ MHz |
| J. Bregman / NASA Ames & D. Rank / Lick Observatory | Photometer/Camera | 2-5, 6-13 | 1/128x128 | Various (Filters) |
| E. Dunham / NASA Ames | High Speed CCD Photometer | 0.3-1.1 | 1/2048x2048 | Various (Filters) |
| E. Erickson / NASA Ames | Echelle Spectrometer | 16-210 | 32/1 | $\lambda/\delta\lambda \sim 1000\text{-}5000$ |
| D. Harper / Yerkes Observatory | Photometer/Camera | 30-500 | 1/8x8 | $\lambda/\delta\lambda \sim 2\text{-}10$ |
| P. Harvey / UT Austin | High Angular Resolution Camera | 40-200 | 1/2x10 | $\lambda/\delta\lambda \sim 20\text{-}100$ |
| T. Herter / Cornell U. | Grating Spectrometer | 5-36 | 128/128 | $\lambda/\delta\lambda \sim 100\text{-}9000$ |
| R. Hildebrand / U. Chicago | Polarimeter | 100 | 1/6x6 | $\lambda/\delta\lambda \sim 2.5$ |
| H. Moseley / NASA GSFC | Grating Spectrometer | 16-150 | 48/1 | $\lambda/\delta\lambda \sim 35\text{-}200$ |
| H. Larson / U. Arizona | Michelson Interferometer | 1-5 | 1 | $\lambda/\delta\lambda \sim 1000\text{-}300{,}000$ |
| H. Röser / DLR Berlin (DE) | Heterodyne Spectrometer | 100-400 | 1400/2 | $\delta\nu \sim 1$ MHz |
| R. Russell / Aerospace Corp. | Prism Spectrometer | 2.9-13.5 | 58/1 & 58/1 | $\lambda/\delta\lambda \sim 25\text{-}120$ |
| G. Stacey / Cornell U. | Imaging Fabry-Perot Spectrometer | 18-42 | 1/128x128 | $\lambda/\delta\lambda \sim 35\text{-}100$ |
| C. Townes / UC Berkeley & R. Genzel / MPE Garching, DE | Imaging Fabry-Perot Spectrometer | 40-200 | 1/5x5 | $\lambda/\delta\lambda \sim 3000\text{-}300{,}000$ |
| F. Witteborn / NASA Ames | Grating Spectrometer | 5-28 | 120/1 | $\lambda/\delta\lambda \sim 300\text{-}1000$ |
| J. Zmuidzinas / CalTech | SIS Heterodyne Spectrometer | 370-600 | 160/1 | $\delta\nu \sim 0.6, 3.0$ MHz |

About 50 specialized science instruments encompassing a wide variety of technologies and capabilities were developed and used by 33 different instrument teams on the KAO during its lifetime. Instrument teams were led by scientists from university, government, and industry laboratories, both U.S. and foreign. They developed the instruments at their home institutions, installed them on the telescope, operated them in flight, and analyzed and published the data. Instrument upgrades were typically made between flight series. The science instruments usually employed the most recently developed or high-tech equipment on the observatory. Probably because they were operated by their developers for their own or for guest investigations, the instruments were actually more reliable than either the aircraft or the telescope system.

A related important if intangible factor in the success of the KAO was the entrepreneurial and enthusiastic spirit it fostered in the investigator teams. Participants were excited by the opportunity – unique in the annals of modern astronomy – to personally prepare for and perform suborbital observations from anywhere on the globe.

The value of this program to the community is evinced in part by the recognitions received by its participants. Some of the awards earned by astronomers experienced with airborne astronomical instrumentation are listed in Table 2. Nine of the sixteen awardees were airborne instrument team leaders. These awards, while not necessarily related directly to research done in the airborne program, demonstrate (1) its appeal for creative application of advanced technologies, and (2) its excellent opportunities for mentoring and developing researchers' skills in observational astronomy and instrumentation. That a majority (four out of seven) of the American Astronomical Society Weber Awards for instrumentation have gone to researchers with extensive airborne astronomy experience attests not only to the effectiveness of the program in fostering opportunities for new instrumentation developments by individual teams, but also to the potential for rapidly advancing infrared and submillimeter technologies.

Table 2. Some Awards Received by Astronomers with Airborne Experience

| Award | Recipients |
|---|---|
| AAS Pierce Prize for outstanding achievement in observational astronomy over the past five years for researchers under 36 years old | Eric E. Becklin[#], Doyal A. Harper[*#], Reinhard Genzel[#], Harriet L. Dinerstein, Kristen Sellgren[*] |
| AAS Cannon Award for outstanding research and promise for future research by a woman within five years of receiving her Ph.D. | Harriet L. Dinerstein, Suzanne Madden |
| AAS Weber Award for Astronomical Instrumentation leading to advances in astronomy | Frank J. Low[#], Thomas G. Phillips[#], Harvey Moseley[*#], James R. Houck[#] |
| ASP Bruce Gold Medal for a lifetime of outstanding research in astronomy | Martin Harwit[#], Frank J. Low[#] |
| ASP Muhlmann Award for innovative advances in astronomical instrumentation | John H. Lacy, Michael Skrutskie |
| Nobel Prize for fundamental work in quantum electronics | Charles H. Townes[#] |
| MacArthur Foundation Award for astrophysics | John E. Carlstrom |
| Pawsey Medal (AU) for excellence in experimental physics | John W. V. Storey |

AAS: American Astronomical Society; ASP: Astronomical Society of the Pacific
[*] indicates Ph.D. thesis included data from airborne observations.
[#] indicates team leader for development of airborne science instrument(s)

Besides the astronomers recognized in Table 2, roughly 200 others – including many graduate students and post-doctoral researchers – participated in the development of instrumentation for airborne observations. The table in the appendix lists some of the scientists whose careers included experience with airborne instrumentation and observations, and who have gone on to make significant contributions in ground- and/or space-based astronomy, including leadership roles in the astronomical community. No matter their subsequent activities, nearly all appreciate and can vouch for the value of their experiences in developing and using airborne instruments. We may expect a substantially larger long-term benefit to the community from the increased instrumentation activity that SOFIA will support.

We recognize that modern focal plane instruments are more complex, expensive, and require longer development periods than those of the KAO era. This is true in all astronomy disciplines. However, the basic merits of SOFIA relative to other facilities for training of personnel and implementation of technology are still valid.

*The Need for Training Instrumentalists*

The development of technically skilled individuals is a national priority. This is made clear in the "America COMPETES Act"[4], a bipartisan congressional response to recommendations contained in the National Academies' "Rising Above the Gathering Storm" report and the Council on Competitiveness' "Innovate America" report. These documents emphasize the need for maintaining and improving innovation in the United States in the 21st Century.

The need is not new. NASA supports astronomy based on the mandate in its charter, the National Aeronautics and Space Act[5], that lists as the agency's first objective: "The expansion of human knowledge of the Earth and of phenomena in the atmosphere and space." Current NASA programs support this mandate, as shown for example by the explicit objective of the Advanced Planning & Integration Office Roadmap[6] "to advance the scientific and technological capabilities of the nation".

To accomplish these objectives requires talented and highly trained personnel. The 2006 National Academy of Science Space Studies Board report *Building a Better NASA Workforce*[7] cites earlier studies of space science and engineering as well as opinions of current experts, to conclude "…there is ultimately no substitute for hands-on training." This principle extends from project managers through systems engineers to specialists skilled with sophisticated astronomical instrumentation and observing techniques, both within and outside of NASA. Explicit technology and training needs for SOFIA and related future space missions are described in the "2008 Community Plan for Far Infrared/Submillimeter Space Astronomy"[8].

*The Need for Infrared/Submillimeter Instrumentation*

Community workshops and studies over the past decade have assessed technology progress and identified and prioritized future needs and the corresponding potential science return. For infrared and submillimeter astronomical research, the most comprehensive is *Detector Needs for Long Wavelength Astrophysics*, A NASA Report by the Infrared, Submillimeter and Millimeter Detector Working Group (ISMDWG, 2002)[9]. Proceedings of the SOFIA 20/20 Vision Workshop (2007)[5] describe subsequent progress and current relevance of the ISMDWG report.

Quotes below are from the executive summary of the latter, with items particularly relevant to SOFIA highlighted in italics:

*"Observations at infrared, submillimeter, and millimeter wavelengths will be essential for addressing many of the key questions in astrophysics.* Because of the very wide wavelength coverage, a variety of detector types will be required to satisfy these needs. To enable and to take full advantage of the opportunities presented by the future mission concepts under consideration, *a significant and diverse effort in developing detector technologies will be needed.*

"The ISMDWG finds that the development of very large ($10^3 – 10^4$ pixels) arrays of direct detectors for far infrared to millimeter wavelengths to be the most important need…. *with the emphasis on producing complete systems.*

*"As detector systems become larger, more complex, and more expensive, the available mechanisms for supporting development from proof of concept to flight worthy technology are limited.* We encourage NASA to develop the resources to support this type of engineering. As part of this finding, *we stress the importance of maintaining key infrastructure elements in the research community…* For coherent systems, the greatest need is improvement in sensitivity between 1 – 3 THz (300 – 100 µm). Additionally, *development in other system components [such as readout technologies and] local oscillators will be needed.* The development of arrays of coherent receivers will greatly increase mapping speed…

*"Continuity and stability of funding is essential to insuring the availability of detectors for future missions."*

Infrared technology is improving rapidly. All of these recommendations are served by the airborne astronomy program, and the test-bed opportunities provided by SOFIA.

### *The Potential of SOFIA*

**Training:** SOFIA will offer the *unique* capability for instrument builders and scientists to make hands-on, real time, astronomical observations with cutting edge technologies at wavelengths obscured from the ground. As studies cited above have found, this is the most effective approach to teaching skills needed for the development and validation of the sophisticated, high technology instrumentation systems that will be required for future space-based observatories.

Valuable lessons and skills for space-mission instrument preparation are taught in the development, deployment and upgrading of science instruments. For and airborne observatory, this is a considerably more structured process than for an average ground based instrument, but is considerably less onerous than that for space instrumentation. While operation on an aircraft makes personnel safety a critical issue, the minutiae of space mission assurance concerns are relatively minimal. Close coordination of all activities – flight planning, instrument airworthiness approval, contingency planning, instrument servicing and maintenance, etc. – is required. Airborne instrument teams must learn to work with a wider range of concerns and staff than ground-based instrument teams.
The airborne instrumentation culture can serve as an effective transitional step between the environments of ground-based and space-based instrumentation.

**Instrumentation:** SOFIA's ongoing investment in new focal plane instrumentation will provide some of the needed resources and continuity for the development and verification of complete instrument systems that have been recommended[9]. As reported in the 20/20 Vision Proceedings[10] "For a significant period of time (after Herschel and before a large space mission like SPICA or SAFIR), SOFIA will be the only [routine] access to much of the key far-infrared and sub-millimeter wavelength range." Balloon-borne platforms can also provide testing of infrared and submillimeter instrumentation, to provide observations such as surveys of the Milky Way and nearby galaxies that can be followed up by SOFIA at higher angular resolution. Both SOFIA and balloons will be valuable for developing instruments incorporating advanced technologies, particularly at Technical Readiness Levels[11] 4 and above. Such test-beds will be efficient and productive means to obtaining the experience needed to reduce risks for the next generation of space observatories.

The needed detector/receiver technologies are still in a relatively primitive state because they have received minimal military or commercial development support. Thus these technologies are appropriate for development at university and government laboratories, as will be encouraged by the availability of and support from SOFIA.

The larger format direct-detection arrays needed for far infrared wavelengths will enable, among numerous other investigations, an efficient census of young stellar objects in nearby (extended) molecular clouds, e.g. Taurus and Ophiuchus, to produce a reasonably complete protostellar classification. As an example, SOFIA's first generation far infrared imager, HAWC[3], currently has a 384 element bolometer detector array. The optics in this camera provide an unvignetted field of view of 6.3 arc minutes diameter. To Nyquist sample this area at a wavelength of 50 µm would require an upgrade to an array of ~25,000 pixels, which would increase the mapping speed by a factor ~60. Such arrays are now foreseeable. Clearly the investment in this technology would be well justified in terms of cost effectiveness on SOFIA, and much more so on future NASA missions. SOFIA can help to bridge the gap to such technology developments.

Similarly, an increase in sensitivity of submillimeter heterodyne receivers would be an extremely valuable addition to NASA's astronomy capabilities. The receivers currently operating at 1-3 THz (300-100 microns) are factors ~10-20 above the fundamental quantum noise limit theoretically achievable. Higher sensitivity could, for example, enable measurement of molecular transitions in cold, prestellar cores to characterize the chemistry affecting the evolution of the earliest stages of star formation. Again, SOFIA offers a test-bed in which cutting edge systems can be exercised while doing high priority science.

In addition to these potential applications, SOFIA will be able to host other significant instrument capabilities[10]. Among these are infrared polarimetry, not available from Herschel or foreseen on JWST, and spectral imaging. Both of these would take advantage of the large format direct detector arrays. In addition, arrays of terahertz heterodyne receivers would vastly improve mapping speeds for high resolution (~1 km/sec) spectroscopy. The related technologies also required to build and evaluate practical, complete instruments will of course be part of these developments.

**Complementarity:** For science, and for development of instrumentalists and novel instrumentation/technology, SOFIA's contributions will complement those of ground-based, other suborbital, and space-borne telescopes.

**Scientific Complementarity:** Scientifically, SOFIA's attributes of world-wide mobility and access to most infrared/submillimeter wavelengths unavailable from the ground assure that airborne and ground-based astronomy are complementary. SOFIA's mobility and flexible scheduling allow rapid deployment for observing ephemeral events (comets, eclipses, occultations, etc) that often escape observations by telescopes in space.

Airborne astronomy is also highly complementary to the infrared and submilimeter *science* capabilities enabled by the other suborbital platforms: sounding rockets and balloons. Balloons and rockets reach higher altitudes than SOFIA and thus offer even higher atmospheric transparency. The long duration balloon program can provide months of time-on-source, which is particularly valuable for surveys. Infrared and submillimeter observations from balloons are thus very valuable scientifically, and can provide results to guide follow-on observations by SOFIA. However, rockets and balloons typically have relatively infrequent launch opportunities and single-purpose science instruments. The observing programs are usually highly focused, and so typically do not support guest investigators.

In contrast, SOFIA will function as a general purpose observatory. It will fly often during the year, offer access to the entire sky, and provide prompt response for observation of targets of opportunity. Its operation and large instrument complement are designed to support guest investigators. Close involvement of astronomers with the science instrument and the flexibility of the platform allows for real time decisions on observational strategies and in dealing with unforeseen contingencies. Clearly SOFIA's science paradigm complements that of balloons and rockets.

Scientific complementarity of SOFIA and space astronomy facilities is assured by NASA's requirement of limited overlap in the capabilities of the concurrent missions it sponsors.

**Instrumentation and Training Complementarity:** Instrumentation development and personnel training for systems operating at wavelengths inaccessible from the ground would of course not be appropriate for ground-based observatories, assuring no overlap with SOFIA. Of course ground-based observatories provide excellent and extensive opportunities for developing talent and instrumentation in their available spectral ranges, and in that sense are akin to SOFIA.

Relative to space-based and other suborbital facilities, SOFIA will offer the *unique* capability of literal hands-on access to its instruments during operation, as well as frequent opportunities for instrument servicing (*e.g.*, cryogen refill), diagnostics, maintenance, upgrades, and exchange. It will afford ample mass, electrical power, and computational infrastructure for the instruments on board. Its instruments and personnel will operate in a shirt-sleeve environment. The access to the instruments removes reliance on telemetry for command and data transfer, and minimizes dependence on remote controlled actuators for adjusting the instrument configuration. These are ideal conditions for training purposes and instrument development focused on the basic performance of the instrument, and assure as well reduced development cost and increased reliability for successful data acquisition.  Thus SOFIA will provide a rich environment for training and creative application of new technology developments, while playing a pivotal role in expanding our understanding of the universe.

Sociologically airborne, balloon and rocket facilities are closely related.  Balloons and rockets fill a valuable role for instrument and personnel development for a broader community than SOFIA, namely one that includes ultraviolet, x-ray, gamma ray, and cosmic ray observational disciplines.  However, the instruments must tolerate the low pressures and temperatures at high altitudes, and must be remotely controlled.  Weight limitations on balloons make it difficult to fly telescopes with apertures as large as SOFIA's, thereby limiting their point source sensitivity and angular resolution.  Instrument support infrastructure (for weight, power, and data processing) is more restrictive. Solving these problems is of value in gaining experience, but must be done in addition to dealing with the intrinsic issues associated with the instrument itself.  Reliability of payload recovery is also concern for balloon and rocket astronomy, especially with regard to high-cost components.  So, while there are similarities among the suborbital facilities, SOFIA remains quite complementary to balloon and rocket programs.

As regards space-borne telescopes, the above discussion makes clear the complementarity between them and SOFIA, which is due largely to the access to the instruments, reflight frequency, platform infrastructure, *etc* provided by SOFIA.

Thus SOFIA can be aptly described as a ground-based observatory that does suborbital astronomy.

## *Conclusion*

Significant, diverse efforts and skilled individuals will be required to develop detector and instrument technologies identified to meet the needs of future NASA astrophysics missions. SOFIA will provide an excellent stimulus, test-bed, and training ground for this work.  With over 100 flights anticipated annually throughout its expected 20 year lifetime, SOFIA will afford frequent and unparalleled opportunities for advancing instrument capabilities and personnel competencies in the field of infrared and submillimeter astronomy.

10. SOFIA 20/20 Vision workshop, 2007: http://www.sofia-vision.caltech.edu/
    www.sofia.usra.edu/Science/07Dec_SOFIA_Vision/SOFIA_2020Vision_white_paper_final.pdf
11. http://en.wikipedia.org/wiki/Technology_Readiness_Level

*Acronyms*

Acronyms used in this paper are given in Table 3.

Table 3. Acronyms

| Acronym | Definition | Acronym | Definition |
|---|---|---|---|
| ALMA | Atacama Large Millimeter Array | KAO | Kuiper Airborne Observatory |
| AU | Australia | Kepler | Satellite to Search for Earthlike Planets |
| CanariCam | Facility mid-infrared imaging spectrograph | LIGO | Laser Interferometer Gravitational Wave Observatory |
| CARA | Center for Astrophysics Research in Antarctica | MIRS | Mid Infrared Spectrometer |
| | | MPE | Max Planck Institute for Extraterrestrial Physics |
| CASIMIR | CAltech Submillimeter Interstellar Medium Investigations Receiver | MPIA | Max Planck Institute for Astronomy |
| | | NICMOS | Near Infrared Camera and Multi Object Spectrometer |
| CEA | Atomic Energy Commission (FR) | | |
| CSO | Caltech Submillimeter Observatory | NL | Netherlands |
| CMBR | Cosmic Microwave Background | PACS | Photodetector Array Camera and Spectrometer |
| DE | Germany | PI | Principal Investigator |
| DLR | German Aerospace Center | PYTHON | CMBR Submillimeter Polarimeter |
| ESO | European Southern Observatory | RIT | Rochester Institute of Technology |
| FIFI LS | Field Imaging Far-Infrared Line Spectrometer | SIS | Superconductor-Insulator-Superconductor |
| | | SAFARI | SpicA FAR-infrared Instrument |
| FORCAST | Faint Object InfraRed CAmera for the SOFIA Telescope | SAFIR | Single Aperture Far Infrared Observatory |
| | | SAFIRE | Submillimeter And Far InfraRed Experiment |
| FR | France | SAO | Smithsonian Astrophysical Observatory |
| HAWC | High-resolution Airborne Wideband Camera | SHARC | Submillimetre High-Angular Resolution Camera |
| | | SHARP | SHARC CII Polarimeter |
| HIFI | Heterodyne Instrument for the Far Infrared | SOFIA | Stratospheric Observatory for Infrared Astronomy |
| | | SPARO | Submillimeter Polarimeter for Antarctic Remote Observing |
| HIPO | High speed Imaging Photometer for Occultations | SPICA | Space Infrared Telescope for Cosmology and Astrophysics (JP) |
| HST | Hubble Space Telescope | | |
| IRAC | Infrared Array Camera | SPIRE | Spectral and Photometric Imaging Receiver |
| IRAS | Infrared Astronomy Satellite | Spitzer | Space Infrared Telescope |
| IRS | Infrared Spectrometer | SRON | Netherlands Institute for Space Research |
| IRTF | Infrared Telescope Facility | STScI | Space Telescope Science Institute |
| IRTS | Infrared Telescope in Space (JP) | SWAS | Submillimeter Wave Astronomy Satellite |
| ISMDWG | Infrared Submillimeter Detector Working Group | T-ReCS | Facility mid-infrared imaging spectrograph |
| | | SWS | Short Wavelength Spectrometer |
| ISO | Infrared Space Observatory | USRA | Universities Space Research Association |
| JWST | James Webb Space Telescope | WISE | Wide-Field Infrared Survey Explorer |
| JP | Japan | | |

*Appendix*

Some Participants in Airborne Instrument Developments and Some Subsequent Contributions

| Scientist | Current Affiliation | Notable Activities |
|---|---|---|
| Eric Becklin* | UCLA/USRA | SOFIA Chief Scientist; former IRTF Director, HST/NICMOS Instrument Team |
| Steve Beckwith* | U. California. | Vice President for Research; former Director, STScI, MPIA |
| John Carlstrom | U. Chicago | Director, Kavli Institute for Cosmological Physics |
| Jackie Davidson* | U. Western Australia | Former SOFIA Project Scientist |
| Jessie Dotson | NASA ARC | SOFIA/HAWC Team, Kepler Science Planning Team |
| Darren Dowell | Caltech | SHARC photometer for CSO |
| Mark Dragovan | JPL/Caltech | CARA /YTHON CMBR Instrument Team |
| Ted Dunham* | Lowell Observatory | PI SOFIA/HIPO; Kepler camera feasibility team |
| Jim Elliot* | MIT | SOFIA/HIPO Team |
| Ed Erickson* | NASA ARC, retired | Original SOFIA Project Scientist for NASA; HST/NICMOS Instrument Team |
| Ian Gatley | RIT | Dean of Science |
| Reinhard Genzel* | MPE, Garching DE | Director, MPE; Herschel /PACS Team |
| Thijs de Graauw* | SRON, Groningen NL | Director, ALMA; PI Herschel/HI-FI, ISO/SWS |
| Matt Greenhouse | NASA GSFC | Project Scientist for JWST Science Instrument Payload |
| D. A. Harper* | U. Chicago | PI SOFIA/HAWC; former director CARA |
| Paul Harvey* | University of Texas | Mission Scientist, Herschel |
| Martin Harwit* | Cornell U., Emeritus | Mission Scientist, Herschel and ISO; SWAS Team |
| Terry Herter* | Cornell University | PI SOFIA/FORCAST; Spitzer support |
| Roger Hildebrand* | U. Chicago, retired | Former Astronomy and Astrophysics Department Chair |
| Jim Houck* | Cornell University | PI Spitzer/IRS; IRAS Co-I |
| Dan Lester | University of Texas | PI for SAFIR Vision Mission Study |
| Frank Low* | Infrared Laboratories | IRAS Co-I, Initial Spitzer Facility Scientist |
| Suzanne Madden | CEA Saclay FR | Herschel/SPIRE, PACS and SPICA/SAFARI instrument teams |
| Gary Melnick | Harvard SAO | PI SWAS, Deputy PI Spitzer/IRAC |
| Alan Moorwood* | ESO | ESO Instrument Program Director |
| Harvey Moseley* | NASA GSFC | PI SOFIA/SAFIRE; detector systems Chandra, JWST |
| Giles Novak | Northwestern U. | Polarimeters SPARO for South Pole; SHARP for CSO |
| Tom Phillips* | Caltech | Director, CSO; U.S. team leader on Herschel |
| Judy Pipher* | U. Rochester, retired | Spitzer/IRAC Team |
| Albrecht Poglitsch | MPE, Garching DE | PI SOFIA/FIFI-LS and Herschel PACS |
| Tom Roellig | NASA ARC | SOFIA Project Scientist for NASA; Spitzer Facility Scientist, IRTS/MIRS (JP) instrument team |
| Hans-Peter Roeser* | U. Stuttgart DE | Managing Director, Institute for Space Systems |
| Michael Skrutskie | U. Virginia | PI, Two Micron All Sky Survey |
| Tom Soifer | Caltech | Director, Spitzer Science Center |
| Charlie Telesco | U. Florida | Project Scientist, T-ReCS on Gemini South, CanariCam on Gran Telescopio Canarias |
| Alan Tokunaga | NASA IRTF Hawaii | Director, NASA IRTF |
| Charles Townes* | UC Berkeley, retired | PI, Ground-based Infrared Spatial Interferometer |
| Mike Werner* | JPL/Caltech | Project Scientist, Spitzer |
| Stan Whitcomb | LIGO/Caltech | Chief Scientist, LIGO |
| Fred Witteborn* | NASA ARC, retired | Original SIRTF (Spitzer) Project Scientist; Kepler camera feasibility team |
| Ned Wright | UCLA | Project Scientist, WISE |
| Jonas Zmuidzinas* | Caltech | PI SOFIA/CASIMIR; Herschel/HIFI instrument team |

* indicates science instrument team leader on the KAO and/or Learjet Observatory